# Extending the class of solvable potentials. IV
## Inverse square potential with a rich spectrum

A. D. Alhaidari

*Saudi Center for Theoretical Physics, P. O. Box 32741, Jeddah 21438, Saudi Arabia*

**Abstract**: This is the fourth article in a series where we succeed in enlarging the class of exactly solvable quantum systems. We do that by working in a complete set of square integrable basis that carries a tridiagonal matrix representation for the wave operator. Consequently, the matrix wave equation becomes a three-term recursion relation for the expansion coefficients of the wavefunction. Finding solutions of this recursion relation in terms of orthogonal polynomials is equivalent to solving the original problem. This method, called the Tridiagonal Representation Approach (TRA), gives a larger class of solvable potentials. Here, we obtain S-wave solutions for a new four-parameter $1/r^2$ singular but short-range radial potential with an elaborate configuration structure and rich spectral property. A particle scattered by this potential has to overcome a barrier then could be trapped within the potential valley in a resonance or bound state. Using complex scaling (complex rotation), we display the rich spectral property of the potential in the complex energy plane for non-zero angular momentum and show how this structure varies with the physical parameters.

## 1. Introduction

In this article, we study the following four-parameter radial potential

$$V(r) = \frac{1}{\sinh^2(\lambda r)}\left[V_0 + V_1 \tanh^2(\lambda r) + V_2 \tanh^4(\lambda r)\right], \tag{1}$$

where $\{\lambda, V_i\}$ are real parameters such that $\lambda > 0$, $V_0 > 0$, and $V_2 \neq 0$. The parameter $\lambda$ has inverse length dimension and it is a measure of the range of the potential. This potential is, in fact, a generalization of the hyperbolic Pöschl-Teller potential [1] as seen by choosing $V_2 = 0$. It has never been studied before. Near the origin, it is $1/r^2$ singular but as $r \to \infty$ it decays exponentially to zero (i.e., it is short-range). There are three distinct physical configurations of this potential as shown in Figure 1. The first one is when the potential has two local extrema (one local minimum and one local maximum). In this configuration, the potential could have resonances but no bound states (Fig. 1a) or it could have both (Fig. 1b). The second configuration occurs when the two extrema coincide at an inflection point (Fig. 1c) or the potential has no local extrema (Fig. 1d). In these two cases, the potential can support neither bound states nor resonances. In the third configuration (Fig. 1e), the potential has one local minimum and could support only bound states but no resonances. Due to the shortness of the potential range, we expect that the size of the bound states energy spectrum to be finite. These speculations will be verified below. Next, we formulate the problem and solve it using the Tridiagonal Representation Approach (TRA) [2-7].

## 2. TRA formulation

We take the system Hamiltonian (in the atomic units $\hbar = m = 1$) as the S-wave second order radial differential operator $H = -\frac{1}{2}\frac{d^2}{dr^2} + V(r)$ and write the wavefunction as the bounded sum $\psi(E,r) = \sum_n f_n(E)\phi_n(r)$, where $\{\phi_n(r)\}$ is a complete set of square integrable basis functions and $\{f_n(E)\}$ are proper expansion coefficients in the energy. If we make the coordinate transformation $x(r) = 2\tanh^2(\lambda r) - 1$, then the Schrödinger wave equation in the new configuration space becomes

$$(H - E)|\psi\rangle = -\frac{1}{2}\left[(x')^2 \frac{d^2}{dx^2} + x'' \frac{d}{dx} - 2V(x) + 2E\right]|\psi\rangle = 0, \qquad (2)$$

where the prime stands for the derivative with respect to $r$. With $x \in [-1,+1]$, we can choose the following square integrable functions in the new configuration space with coordinate $x$ as basis elements for the expansion of the wavefunction

$$\phi_n(x) = A_n(1-x)^\alpha(1+x)^\beta P_n^{(\mu,\nu)}(x), \qquad (3)$$

where $P_n^{(\mu,\nu)}(x)$ is the Jacobi polynomial of degree $n = 0,1,2,..$ in $x$. The parameters $\mu$ and $\nu$ are larger than $-1$ whereas the boundary conditions and square integrability (with respect to the integral measure $dx$) require that $x'(1-x)^{2\alpha-1}(1+x)^{2\beta-1} = (1-x)^\mu(1+x)^\nu$ giving $2\alpha = \mu$ and $2\beta = \nu + \frac{1}{2}$ since $x' = \sqrt{2}\lambda(1-x)\sqrt{1+x}$. We choose the normalization constant as $A_n = \sqrt{\frac{2n+\mu+\nu+1}{2^{\mu+\nu+\frac{1}{2}}} \frac{\Gamma(n+1)\Gamma(n+\mu+\nu+1)}{\Gamma(n+\nu+1)\Gamma(n+\mu+1)}}$. From Eq. (2), we can write the wave operator $J = H - E$ as

$$J = -\lambda^2(1-x)\left[(1-x^2)\frac{d^2}{dx^2} - \frac{1+3x}{2}\frac{d}{dx} + \frac{\varepsilon}{1-x} - \frac{u_1}{2} - \frac{u_0}{1+x} - \frac{u_2}{4}(1+x)\right], \qquad (4)$$

where $\varepsilon = E/\lambda^2$, $u_i = V_i/\lambda^2$ and we have used $x''/(x')^2 = -\frac{1}{2}\frac{1+3x}{1-x^2}$ and wrote the potential function in the new coordinate $x$ as $V(x) = (1-x)\left[\frac{V_1}{2} + \frac{V_0}{1+x} + \frac{V_2}{4}(1+x)\right]$. Consequently, the action of the wave operator on the basis elements (3) is calculated as

$$J|\phi_n\rangle = -\lambda^2 A_n(1-x)^{1+\frac{\mu}{2}}(1+x)^{\frac{1}{4}+\frac{\nu}{2}} \times$$
$$\left[\frac{1}{2}\frac{2\varepsilon+\mu^2}{1-x} - \frac{1}{2}\frac{2u_0 - \nu^2 + \frac{1}{4}}{1+x} - \frac{u_2}{4}(1+x) - \left(n + \frac{\mu+\nu+1}{2}\right)^2 + \frac{1}{16} - \frac{u_1}{2}\right]P_n^{(\mu,\nu)}(x) \qquad (5)$$

where we have used the differential equation of the Jacobi polynomial that reads $(1-x^2)\frac{d^2}{dx^2}P_n^{(\mu,\nu)} - \left[(\mu+\nu+2)x + \mu - \nu\right]\frac{d}{dx}P_n^{(\mu,\nu)} + n(n+\mu+\nu+1)P_n^{(\mu,\nu)} = 0$. Thus, the matrix elements of the wave operator becomes

$$J_{m,n} = \langle\phi_m|J|\phi_n\rangle = -\lambda^2 \frac{A_m A_n}{\sqrt{2}} \int_{-1}^{+1}(1-x)^\mu(1+x)^\nu F(x) P_m^{(\mu,\nu)}(x) P_n^{(\mu,\nu)}(x) dx, \qquad (6)$$

where $F(x)$ is the expression inside the square bracket in Eq. (5) and we have used the integral transform $\int_0^\infty \ldots dr = \int_{-1}^{+1} \ldots \frac{dx}{x'}$.

Now, in the TRA, the matrix representation of the wave operator (6) is required to be tridiagonal and symmetric. The recursion relation of the Jacobi polynomial and its orthogonality [8] imply that this requirement is satisfied if and only if the function $F(x)$ in Eq. (6) is linear in $x$. Therefore, to eliminate the two non-linear terms $\frac{1}{1\pm x}$ in $F(x)$ the basis parameters must be chosen as follows

$$\mu^2 = -2\varepsilon \text{ and } \nu^2 = 2u_0 + \tfrac{1}{4}. \tag{7}$$

This implies that the solution is restricted to negative energies and that $u_0 > -\tfrac{1}{8}$. The latter condition is automatically satisfied since $V_0 > 0$. Negative energy solutions correspond to bound states and dictate that the potential configuration be either the one shown in Fig. 1b or that shown in Fig. 1e. The constraints (7) together with the three-term recursion relation of the Jacobi polynomials and their orthogonality property, $\frac{1}{\sqrt{2}} A_n^2 \int_{-1}^{+1} (1-x)^\mu (1-x)^\nu P_n^{(\mu,\nu)}(x) P_m^{(\mu,\nu)}(x) dx = \delta_{n,m}$, give the following tridiagonal and symmetric matrix representation for the wave operator (6)

$$\frac{1}{\lambda^2} J_{n,m} = \left[\left(n + \tfrac{\mu+\nu+1}{2}\right)^2 - \tfrac{1}{16} + \tfrac{u_1}{2} + \tfrac{u_2}{4}(1+C_n)\right]\delta_{n,m} + \tfrac{u_2}{4}\left(D_{n-1}\delta_{n,m+1} + D_n\delta_{n,m-1}\right), \tag{8}$$

where $C_n = \frac{\nu^2 - \mu^2}{(2n+\mu+\nu)(2n+\mu+\nu+2)}$ and $D_n = \frac{2}{2n+\mu+\nu+2}\sqrt{\frac{(n+1)(n+\mu+1)(n+\nu+1)(n+\mu+\nu+1)}{(2n+\mu+\nu+1)(2n+\mu+\nu+3)}}$.

Thus, the matrix wave equation $\langle \phi_n | J | \psi \rangle = \sum_m J_{n,m} f_m = 0$ gives the following symmetric three-term recursion relation for the expansion coefficients of the wave function

$$\left(\tfrac{1}{4} - 2u_1\right) P_n = \left[(2n+\mu+\nu+1)^2 + u_2(1+C_n)\right] P_n + u_2\left(D_{n-1}P_{n-1} + D_n P_{n+1}\right), \tag{9}$$

where we wrote $f_n(E) = f_0(E) P_n(\varepsilon)$ making $P_0 = 1$. This relation is valid for $n = 1, 2, 3, \ldots$ with $P_0 = 1$ and $P_1 = \frac{-1}{u_2 D_0}\left[2u_1 + 2u_2 \frac{\nu+1}{\mu+\nu+2} + (\mu+\nu+1)^2 - \tfrac{1}{4}\right]$. This recursion relation gives $P_n$ as a polynomial in $\left(\tfrac{1}{4} - 2u_1\right)$ of degree $n$. This polynomial is not found in the appropriate mathematics literature. Its analytic properties that include the weight function, generating function, orthogonality, zeros, etc. are yet to be derived. However, this polynomial has been encountered frequently in the physics literature while solving various problems in quantum mechanics [2-7]. Had, for example, the asymptotics of the polynomial ($\lim_{n\to\infty} P_n$) been known we could have simply read off the energy spectrum $\{\varepsilon_m\}$ from the condition that makes this asymptotics vanish at these energies. This same condition gives the discrete version of the polynomial $P_n(\varepsilon_m)$ that enters as expansion coefficients of the corresponding bound state wavefunction $\psi(E_m, r)$. In the absence of knowledge of the analytic properties of these polynomials including their asymptotics, we follow in the next section a procedure called the "*potential parameter spectrum*" (PPS) [5,6,9] to obtain the energy spectrum and the corresponding wavefunctions.

Note that if $V_2 = 0$, then the potential (1) becomes the well-known hyperbolic Pöschl-Teller potential, which belongs to the conventional class of exactly solvable potentials in the standard formulation of quantum mechanics [1]. In this case, the tridiagonal representation reduces to a diagonal one and the energy spectrum is obtained from (8) or (9) that read $\left(\frac{1}{4} - 2u_1\right) = \left(2n + \mu + \nu + 1\right)^2$ giving the well-known energy spectrum formula

$$\varepsilon_n = -\frac{1}{2}\left(2n + 1 + \sqrt{\frac{1}{4} + 2u_0} - \sqrt{\frac{1}{4} - 2u_1}\right)^2, \tag{10}$$

where $n = 0, 1, 2, ..., n_{max}$ and $n_{max}$ is the largest integer less than or equal to $\frac{1}{2}\left|1 + \sqrt{\frac{1}{4} + 2u_0} - \sqrt{\frac{1}{4} - 2u_1}\right|$.

## 3. PPS procedure

The procedure goes as follows: We write the recursion relation (9) as the eigenvalue matrix equation $\Sigma|P\rangle = \left(\frac{1}{4} - 2u_1\right)|P\rangle$, where $\Sigma$ is the tridiagonal symmetric matrix

$$\Sigma_{n,m} = \left[\left(2n + \mu + \nu + 1\right)^2 + u_2\left(1 + C_n\right)\right]\delta_{n,m} + u_2\left(D_{n-1}\delta_{n,m+1} + D_n\delta_{n,m-1}\right). \tag{11}$$

This is an energy dependent matrix via the parameter $\mu$. Therefore, for a given negative energy and fixed parameters $(u_0, u_2)$, the (infinite) eigenvalues of this matrix $\{\frac{1}{4} - 2u_1\}$ correspond to all potential functions with the (infinite) set of values of the parameter $\{u_1\}$ and $(u_0, u_2)$ that solve the wave equation at that given energy. Figure 2a shows the lowest discrete parameter spectrum of $u_1$ for a given range of negative energies and for fixed parameters $(u_0, u_2)$. Therefore, a horizontal line in the figure at $u_1 = w$ as shown in Fig. 2b intersects the curves at the energy spectrum for the potential (1) with the parameters $(V_0, V_1, V_2) = \lambda^2(u_0, w, u_2)$. It is then obvious from Fig. 2 that the energy spectrum is finite since the horizontal line intersects the curves at a finite number of points. The number of these points, which is the size of the energy spectrum, increases as the horizontal line is lowered (i.e., as $w$ increases in negative value and the potential valley becomes deeper). We should note that in the PPS computations, we use any appropriate numerical fitting routine for the $\{u_1\}$ eigenvalues on a single $m^{th}$ curve in Fig. 2a giving $u_1(\varepsilon_m)$, where $m = 0, 1, 2, ..$ from top to bottom in the figure. Subsequently, we invert these functions to obtain $\{\varepsilon_m(u_1)\}$, which is the energy spectrum of the potential (1) with the parameters $(u_0, u_1, u_2)$.

Table 1 is a list of the finite energy spectrum obtained by the PPS procedure for a given set of values of the potential parameters $(\lambda, V_0, V_1, V_2)$ and for various basis sizes. With these parameters, the potential configuration is similar to that shown in Fig. 1e. The Table shows clearly the rapid convergence of these values with the size of the basis, which is a unique property of the PPS procedure. In the calculation, we used the continued fraction fitting routine based on the rational fraction approximation of Haymaker and Schlessinger similar to that in the Padé method [10]. To demonstrate the exceptional accuracy of the PPS results, we compare them to those obtained by direct numerical diagonalization of the Hamiltonian matrix with the same basis size. The Hamiltonian matrix is obtained from Eq. (8) as $H = J\big|_{E=0=\mu}$. Then, the energy spectrum is calculated from the wave equation $H|\psi\rangle = E|\psi\rangle$ as the generalized eigenvalues $\{E\}$ of the matrix equation $\sum_m H_{n,m} f_m = E \sum_m \Omega_{n,m} f_m$, where $\Omega_{n,m} = \langle \phi_n | \phi_m \rangle$. Table 2 shows a comparison of the PPS results with those of the Hamiltonian diagonalization method (HD) for the same basis size of 50. We show only significant decimal digits that do not change with any substantial increase in the basis size (e.g. from size 50 to 60). It is obvious that PPS has a superior accuracy. In the following section, we will also have an independent confirmation of this observation by employing another method for calculating not just the bound states but also resonance energies called the Complex Scaling (CS). The result of that calculation, which will be detailed below, for a basis size of 50 is given as the third column of Table 2 with the heading CS. Again, we keep only significant decimal places. Although, CS is more accurate than the HD, it is still less accurate than the PPS as indicated by the value of the highest bound state energy.

In Figure 3, we plot the bound state wavefunctions corresponding to the physical configuration and energy spectrum of Table 1. We calculate the $m^{th}$ bound state using the sum $\psi(E_m, r) \sim \sum_{n=0}^{N-1} P_n(\varepsilon_m) \phi_n(r)$, where $N$ is some large enough integer. We note that as $N$ increases from small values, the plot becomes stable for a range of values of $N$. Then, as $N$ increases beyond some critical value the sum starts becoming unstable and produce only oscillations that increases in number and magnitude. This critical value increases with the energy level $m$. For Figure 3, these critical values for the four bound states are in the range 6 to 9. Moreover, if we try to evaluate the sum at an energy other than those of the bound states then we will never reach stable results but only oscillations that increases in number and magnitude.

## 4. Spectral property of the potential

The exact TRA solution obtained above for the potential $V(r)$ of Eq. (1) was only for bound states and for zero angular momentum. However, for non-zero angular momentum, we can use the complex scaling method (a.k.a. complex rotation method [11-21]) to obtain a highly accurate evaluation of both bound states and resonance energies. Now, the expansion of the potential (1) near the origin is

$$\lim_{r \to 0} V(r) \approx \frac{V_0/\lambda^2}{r^2} + \left(V_1 - \frac{V_0}{3}\right) + (\lambda r)^2 \left(V_2 - V_1 + \frac{V_0}{15}\right). \tag{12}$$

Thus, writing the potential as $V(r) = \frac{u_0}{r^2} + \tilde{V}(r)$ makes $\tilde{V}(r)$ a non-singular short-range potential and gives the total Hamiltonian as

$$H = -\frac{1}{2}\frac{d^2}{dr^2} + \frac{\ell(\ell+1) + 2u_0}{2r^2} + \tilde{V}(r). \tag{13}$$

Since $u_0 > 0$, then this corresponds to a three-dimensional system with spherical symmetry, short-range non-singular potential $\tilde{V}(r)$ and with an effective non-integral angular momentum $\tilde{\ell} = -\frac{1}{2} + \sqrt{\left(\ell + \frac{1}{2}\right)^2 + 2u_0}$. Consequently, the system is well suited for treatment by the complex scaling method. To perform the CS calculation, we use the following square integrable Laguerre basis

$$\chi_n(r) = \sqrt{\frac{\Gamma(n+1)}{\Gamma(n+2\tilde{\ell}+2)}} (\gamma r)^{\tilde{\ell}+1} e^{-\gamma r/2} L_n^{2\tilde{\ell}+1}(\gamma r), \tag{14}$$

where $L_n^v(z)$ is the Laguerre polynomial of degree $n$ in $z$ and $\gamma^{-1}$ is a length scale computational parameter. In this basis, the representation of the kinetic energy part of the Hamiltonian (13) becomes the following tridiagonal symmetric matrix

$$\frac{1}{2}\langle \chi_n | \left[ -\frac{d^2}{dr^2} + \frac{\tilde{\ell}(\tilde{\ell}+1)}{r^2} \right] | \chi_m \rangle = \frac{\gamma^2}{4}\left[ (n+\tilde{\ell}+1)\delta_{n,m} + \tfrac{1}{2}\sqrt{n(n+2\tilde{\ell}+1)}\delta_{n,m+1} + \tfrac{1}{2}\sqrt{(n+1)(n+2\tilde{\ell}+2)}\delta_{n,m-1} \right] \tag{15}$$

To calculate the matrix elements of the potential $\langle \chi_n | \tilde{V} | \chi_m \rangle$, we use Gauss quadrature integral approximation associated with the Laguerre polynomials. Due to the shortness of the potential range and its non-singularity, the calculation of the matrix elements becomes highly accurate. Adding the matrix elements of the kinetic energy to the potential gives the total Hamiltonian matrix. Complex scaling is accomplished by choosing the length scale computational parameter as $\gamma = \rho e^{-i\theta}$, where $\rho$ and $\theta$ are real non-physical parameters and $\theta < \tfrac{1}{2}\pi$. Finally, we obtain the complex (Harris) eigenvalues of the Hamiltonian using the wave equation $H|\psi\rangle = E|\psi\rangle$ as the generalized eigenvalues of the matrix equation $\sum_{m=0}^{N-1} H_{n,m} f_m = E \sum_{m=0}^{N-1} \Omega_{n,m} f_m$, where $N$ is the size of the basis $\{\chi_n\}_{n=0}^{N-1}$ and

$$\Omega_{n,m} = \langle \chi_n | \chi_m \rangle = 2(n+\tilde{\ell}+1)\delta_{n,m} - \sqrt{n(n+2\tilde{\ell}+1)}\,\delta_{n,m+1} - \sqrt{(n+1)(n+2\tilde{\ell}+2)}\,\delta_{n,m-1}. \tag{16}$$

Now, out of the $N$ Harris eigenvalues in the complex energy plane the physically relevant ones are those that do not change (within a chosen accuracy) with variations in the computational parameters $\rho$ and $\theta$. Those eigenvalues split into two sets. The ones that are located on the negative real energy axis correspond to bound states whereas the ones that are located in the lower half of the complex energy plane between the positive real line and the ray $-2\theta$ correspond to resonances.

Table 3 is a list of bound state and resonance energies for a given set of potential parameters and for various values of the angular momentum. With these parameter values, the potential configuration is similar to that shown in Fig. 1b. The basis size was taken 50. Figure 4 gives the spectral diagram of the potential function (1) in the complex energy plane for the same values of the physical parameters as in Table 3. In the diagram, squares (circles) represent bound (resonance) state energies, respectively. Additionally, the video animation shows how the bound states and resonance energies move in the complex energy plane as $V_1$ varies while keeping the rest of the potential parameters fixed. As $V_1$ increases (the bottom of the potential valley rises) bound states move to the right (i.e., energy levels increase) on the real energy line until they cross the origin then move slowly into the lower half of the complex energy plane turning into resonances. Therefore, bound states are reduced by one while resonances gain an extra one. Moreover, the animation shows that resonances move diagonally (southeast) in the energy plane exposing resonances that were hidden in the lower half of the complex energy plane between the negative real line and the $-2\theta$ ray.

## 5. Conclusion

Using the tridiagonal representation approach, we obtained an exact bound states S-wave solution for the radial potential (1), which was not studied before in the physics literature and does not belong to the conventional class of exactly solvable problems. This potential can support bound states and/or resonances in three different configurations. Using the complex scaling method, we showed the rich structure of its bound states and resonance spectrum. We also showed how that spectrum changes with variations in the physical parameters. One very important issue yet to be resolved is to derive the analytic properties of the orthogonal polynomial that satisfies the recursion relation (9). This is a pure mathematical task to be taken up, hopefully, by specialists in the field.

## Tables Captions

**Table 1**: The finite energy spectrum in units of $-\lambda^2$ obtained by the PPS procedure for the potential parameters $(u_0, u_1, u_2) = (1, -50, 2)$ and for various basis sizes. Rapid convergence with the basis size is obvious.

**Table 2**: Comparison of the PPS results of Table 1 with those obtained by the Hamiltonian diagonalization (HD) and complex scaling (CS) methods for the same basis size of 50. Only significant digits are kept.

**Table 3**: Bound states and resonances energies in units of $\lambda^2$ for the potential parameters $(u_0, u_1, u_2) = (2, -80, 120)$. We took the non-physical computational parameters as $N = 50$, $\theta$ equals to 0.0 and 0.8 radians for bound states and resonances, respectively. Whereas, we took $\rho$ as shown in the Table. Again, only significant digits are kept.

## Figures Captions

**Fig. 1**: The distinct configurations of the potential that could support: (a) only resonances, (b) both resonances and bound states, (c) & (d) none, and (e) only bound states.

**Fig. 2**: (a) The lowest discrete spectrum of the potential parameter $u_1$ for a given range of negative energies and fixed parameters $(u_0, u_2) = (1.0, 2.0)$. (b) The finite bound states energy spectrum of the potential with the parameters $(V_0, V_1, V_2) = \lambda^2 (u_0, w, u_2)$ is obtained as the intersection of the horizontal line $u_1 = w$ with the curves.

**Fig. 3**: The bound state wavefunctions corresponding to the physical configuration and four energy eigenvalues of Table 1. The radial coordinate is measured in unit of $\lambda^{-1}$.

**Fig. 4**: The bound states (squares) and resonance (circles) energies in the complex energy plane for the potential with the parameters $(u_0, u_1, u_2) = (2, -80, 120)$ and for various values of the angular momentum quantum number $\ell$. We took the computational parameters as $N = 100$, $\theta = 0.8$ radians and $\rho$ as shown in the Table 3.

## Animation Caption

Video animation showing how the bound states and resonance energies move in the complex energy plane as $V_1$ varies from $-100\lambda^2$ to $-40\lambda^2$ (as displayed in the video) while keeping the rest of the potential parameters fixed at $(V_0, V_2) = \lambda^2 (2, 120)$ and with $\ell = 0$. We took the computational parameters as $N = 100$, $2\theta = \frac{7}{4}$ radians and $\rho = 40\lambda$.

**Table 1**

| n | 4×4 | 6×6 | 10×10 | 100×100 |
|---|---|---|---|---|
| 0 | 27.878950096075 | 27.878950096074 | 27.878950096074 | 27.878950096074 |
| 1 | 14.799140053549 | 14.799140053574 | 14.799140053574 | 14.799140053574 |
| 2 | 5.854540858323 | 5.854541479288 | 5.854541479288 | 5.854541479288 |
| 3 | 0.994844848888 | 0.996376819202 | 0.996376819225 | 0.996376819225 |

**Table 2**

| n | PPS | HD | CS |
|---|---|---|---|
| 0 | 27.878950096074 | 27.878950096074 | 27.878950096074 |
| 1 | 14.799140053574 | 14.79914005357 | 14.799140053574 |
| 2 | 5.854541479288 | 5.85454148 | 5.854541479288 |
| 3 | 0.996376819225 | 0.9967 | 0.996376819 |

**Table 3**

| $\ell$ | Bound States | Resonances |
|---|---|---|
| 0 | −27.66703017245<br>−4.96995355885<br>($\rho = 40\lambda$) | 5.1432  −i 1.73656<br>5.7767  −i 12.3187<br>1.61    −i 29.27<br>($\rho = 40\lambda$) |
| 1 | −21.21593606495<br>−0.8517865495<br>($\rho = 25\lambda$) | 6.2706  −i 3.4478<br>6.038   −i 15.8152<br>1.154   −i 33.87<br>($\rho = 40\lambda$) |
| 2 | −11.585302647445<br>($\rho = 50\lambda$) | 4.3251234  −i 0.244407<br>7.998469   −i 7.512996<br>6.5784     −i 22.0054<br>0.53       −i 41.6<br>($\rho = 50\lambda$) |
| 3 | −1.44701935596<br>($\rho = 30\lambda$) | 8.59697 −i 2.2622<br>10.2802 −i 13.407<br>7.414   −i 29.9473<br>($\rho = 35\lambda$) |

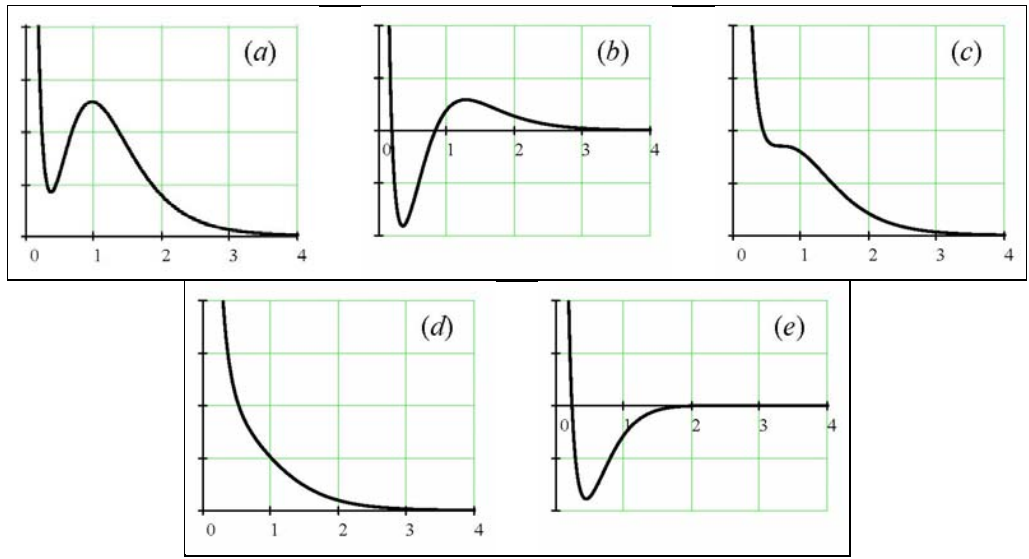

**Fig. 1**

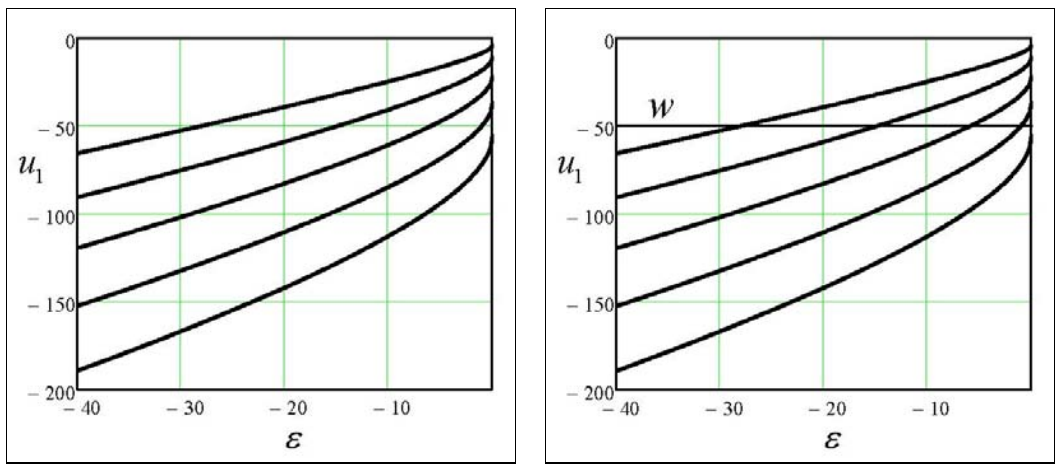

**Fig. 2**

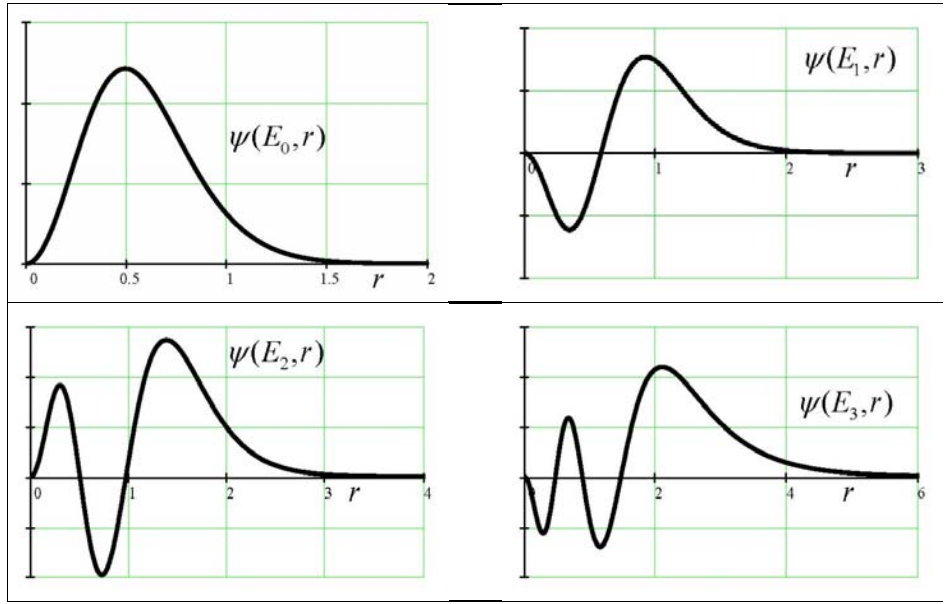

**Fig. 3**

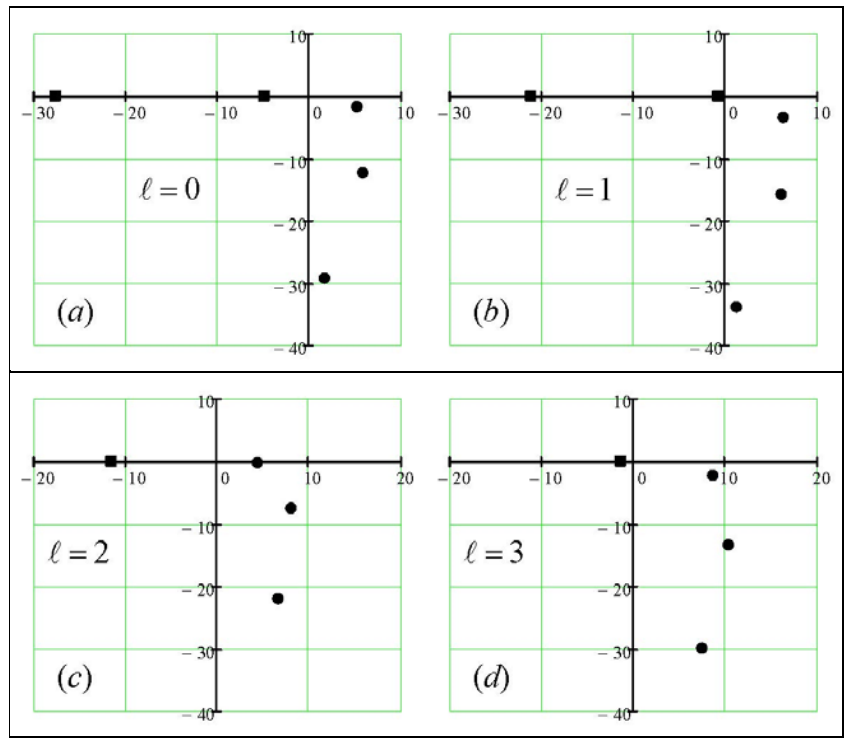

**Fig. 4**